\documentclass[12pt]{article}

\usepackage{amsmath,amssymb}
\usepackage{graphicx}

\makeatletter

\renewcommand\section{\@startsection{section}{1}{\z@}
                                   {-3.5ex \@plus -1ex \@minus -.2ex}
                                   {2.3ex \@plus .2ex}
                                   {\normalfont\large\bfseries}}
\renewcommand\subsection{\@startsection{subsection}{2}{\z@}
                                   {-3.25ex\@plus -1ex \@minus -.2ex}
                                   {1.5ex \@plus .2ex}
                                   {\normalfont\normalsize\bfseries}}
\renewcommand\subsubsection{\@startsection{subsubsection}{3}{\z@}
                                   {-3.25ex\@plus -1ex \@minus -.2ex}
                                   {1.5ex \@plus .2ex}
                                   {\normalfont\normalsize\bfseries}}
\renewcommand\paragraph{\@startsection{paragraph}{4}{\z@}
                                   {3.25ex \@plus1ex \@minus.2ex}
                                   {-1em}
                                   {\normalfont\normalsize\bfseries}}

\makeatother

\newcommand{\beq}{\begin{equation}}
\newcommand{\eeq}{\end{equation}}
\newcommand{\bea}{\begin{eqnarray}}
\newcommand{\eea}{\end{eqnarray}}
\newcommand{\SL}{\rm SL}
\newcommand{\SU}{\rm SU}
\newcommand{\SO}{\rm SO}
\newcommand{\Sp}{\rm Sp}
\newcommand{\Spin}{\rm Spin}
\newcommand{\su}{\rm su}
\newcommand{\so}{\rm so}
\newcommand{\symp}{\rm sp}
\newcommand{\Z}{\mathbb Z}
\newcommand{\Pe}{P_{\rm even}}
\newcommand{\Po}{P_{\rm odd}}
\newcommand{\id}{\hbox{1\kern-.27em l}}
\newcommand{\Gadj}{G_{\rm adj}}
\newcommand {\Pf}{\rm Pf}
\newcommand{\column}[4]{\left( \begin{array}{c} #1 \cr #2 \cr #3 \cr #4 \end{array} \right)}

\begin{document}

\pagestyle{empty}

\begin{center}

\vspace*{30mm}
{\Large  The low-energy spectrum of $(2, 0)$ theory on $T^5 \times \mathbb R$}

\vspace*{30mm}
{\large M{\aa}ns Henningson}

\vspace*{5mm}
Department of Fundamental Physics\\
Chalmers University of Technology\\
S-412 96 G\"oteborg, Sweden\\[3mm]
{\tt mans@chalmers.se}     
     
\vspace*{30mm}{\bf Abstract:} 
\end{center}
We consider the $ADE$-series of $(2, 0)$ supersymmetric quantum theories on $T^5 \times \mathbb R$, where the first factor is a flat spatial five-torus, and the second factor denotes time. The quantum states of such a theory $\Phi$ are characterized by a discrete quantum number $f \in H^3 (T^5, C)$, where the finite abelian group $C$ is the center subgroup of the corresponding simply connected simply laced Lie group $G$.  At energies that are low compared to the inverse size of the $T^5$, the spectrum consists of a set of continua of states, each of which is characterized by the value of $f$ and some number  $5 r$ of additional continuous parameters. By exploiting the interpretation of this theory as the ultraviolet completion of maximally supersymmetric Yang-Mills theory  on $T^4 \times S^1 \times \mathbb R$ with gauge group $G_{\rm adj} = G / C$ and coupling constant $g$ given by the square root of the radius of the $S^1$ factor, one may compute the number $N_f^r (\Phi)$ of such continua. We perform these calculations in detail for the $A$- and $D$-series. While the Yang-Mills theory formalism is manifestly invariant under the $\SL_4 (\Z)$ mapping class group of $T^4$, the results are actually found to be invariant under the $\SL_5 (\Z)$ mapping class group of $T^5$, which provides a strong consistency check.

\newpage

\pagestyle{plain}

\setcounter{equation}{0}
\section{Introduction}
Understanding the conceptual foundations of the $ADE$-series of $(2, 0)$ supersymmetric quantum theories in $d = 5 + 1$ dimensions \cite{Witten95} remains an outstanding challenge. In this paper, we will consider these theories on a space-time of the form 
\beq
T^5 \times \mathbb R , 
\eeq 
where the first factor is a flat spatial five-torus, and the second factor denotes time. By supersymmetry, the energy spectrum of such a system is bounded from below by zero. In general, the precise spectrum depends on the choice of flat metric on $T^5$, and a complete determination of it is certainly out of reach at the present. But the properties of the low-energy spectrum (compared to the scale set by the inverse size of the $T^5$) are independent of the $T^5$ geometry, and should be more accessible. The theory has no mass-gap, so the low-energy spectrum consists of a set of continua of states, each of which is characterized by the values of some discrete quantum numbers, and also by the number of continuous parameters needed to label the states. The goal of this paper is to describe the properties of this low-energy spectrum in general, and also to compute it explicitly in most cases.

What makes this problem tractable is the relationship between $(2, 0)$ theory in $d = 5 + 1$ dimensions and maximally supersymmetric Yang-Mills theory in lower dimensions \cite{Witten02}. In particular, with
\beq
T^5 \simeq T^4 \times S^1 ,
\eeq
our $(2, 0)$ theory $\Phi$ can be regarded as an ultra-violet completion of maximally supersymmetric Yang-Mills theory on $T^4 \times \mathbb R$.  The gauge group of the Yang-Mills theory is given by the simply laced group of adjoint type $\Gadj$ whose root lattice $\Gamma_{\rm root}$ is determined by $\Phi$. We have $\Gadj = G / C$, where $G$ is the simply connected covering group of $\Gadj$ with center subgroup $C$.  In terms of lattices, $C \simeq \Gamma_{\rm weight} / \Gamma_{\rm root}$, where the weight lattice $\Gamma_{\rm weight}$ is the dual of the root lattice $\Gamma_{\rm root} \subset \Gamma_{\rm weight}$. The Yang-Mills coupling constant $g$ is given by the square root of the radius of the $S^1$ factor. The independence of the low-energy spectrum of the $T^5$ geometry in particular means that it is independent of the coupling constant, and thus may be computed by semi-classical methods in the weak coupling limit $g \rightarrow 0$. This is completely analogous to previous calculations for maximally supersymmetric Yang-Mills theory on $T^3 \times \mathbb R$ \cite{Henningson-Wyllard}.

In the next section, we will outline the general features of this computation. In sections three and four, we will then perform it in detail for the $A$-series and the $D$-series respectively. The remaining three models of $E$-type are left for the future. From the point of view of $(2, 0)$ theory, these results derived in Yang-Mills theory, may be regarded as "experimental" data. Hopefully, they will eventually get a more "theoretical" explanation, shedding light on the conceptual foundations of $(2, 0)$ theory.

\setcounter{equation}{0}
\section{The general theory}
In this section, we will review maximally supersymmetric Yang-Mills theory with a simply laced gauge group $\Gadj = G / C$ of adjoint type on $T^4 \times \mathbb R$ and explain how to determine the low-energy spectrum. We will also describe the interpretation of the results from the perspective of $(2, 0)$ theory.

\subsection{The degrees of freedom}
We should first determine the possible topological choices of the Yang-Mills gauge bundle $P$ (a principal $\Gadj$ bundle over the spatial $T^4$). The relevant homotopy groups of the gauge group $\Gadj = G / C$, where $G$ is simply connected with center subgroup $C$, are 
\beq
 \pi_k (\Gadj) \simeq \left\{
 \begin{array}{ll}
 1, & k = 0 \cr
 C, & k = 1 \cr
 1, & k = 2 \cr
 \mathbb Z & k = 3 .
 \end{array}
 \right.
 \eeq
It follows that the isomorphism class of $P$  is completely determined by two characteristic classes: The discrete abelian magnetic 't~Hooft flux \cite{tHooft}
\beq
m \in H^2 (T^4, C) ,
\eeq
and the (fractional) instanton number
\beq
k \in H^4 (T^4, \mathbb Q) \simeq H^0 (T^4, \mathbb Q) \simeq \mathbb Q .
\eeq
These are related as
\beq \label{fraction}
k - \frac{1}{2} m \cdot m \in H^4 (T^4, \Z) \subset H^4 (T^4, \mathbb Q) , 
\eeq
where the raised dot denotes the tensor product of the cup product on $H^2 (T^4, \Z)$ and the inner product modulo integers on $C$. (The inner product on $C \simeq \Gamma_{\rm weight} / \Gamma_{\rm root}$ is induced from the inner product on $\Gamma_{\rm weight}$.)

For a given bundle $P$, we let $\tilde{\Omega} = {\rm Aut} (P)$ denote the group of gauge transformations (bundle automorphisms). It is parametrized by the space of sections of the bundle ${\rm Ad} (P)$ associated to $P$ via the adjoint action of $G_{\rm adj}$ on itself. In general, $\tilde{\Omega}$ is disconnected, and we let $\Omega_0$ denote its connected component subgroup. Physical states must be invariant under $\Omega_0$. Indeed, the generators of infinitesimal gauge transformations are weakly equal to zero in the classical field theory. But the states may transform non-trivially under the discrete abelian quotient group $\Omega = \tilde{\Omega} / \Omega_0$ of "large" gauge transformations. It follows from the homotopy type of $\Gadj$ that
\beq
\Omega \simeq {\rm Hom} (\pi_1 (T^4), \pi_1 (\Gadj)) \simeq H^1 (T^4, C) .
\eeq
The transformation properties of a quantum state can thus be described by the discrete abelian electric 't~Hooft flux \cite{tHooft}
\beq
e \in {\rm Hom} (\Omega, U (1)) \simeq H^3 (T^4, C^*) \simeq H^3 (T^4, C) ,
\eeq
where we have first used Poincar\'e duality and then the isomorphism between $C$ and its Pontryagin dual $C^* = {\rm Hom} (C, U(1))$ that follows from the inner product on $C$.

We will work in temporal gauge, so that the time-component of the gauge field is identically zero. The bosonic fields of the theory are then a connection $A$ on the bundle $P$ over $T^4$, and five scalar fields $\Phi^5, \ldots, \Phi^9$ that are sections of the vector bundle ${\rm ad} (P)$ associated to $P$ via the adjoint representation of $\Gadj$.  The scalar fields transform in the five-dimensional vector representation of the $\Spin (5) \simeq \Sp (4)$ $R$-symmetry group. The fermionic fields are four spinors fields $\Psi^1, \ldots, \Psi^4$ that are sections of ${\rm ad} (P) \otimes S$, where $S$ is a spinor bundle over space-time. They transform in the four-dimensional spinor representation of $\Spin (5) \simeq \Sp (4)$.

\subsection{Low energy states}
The Yang-Mills energy density is a sum of positive definite terms, each of which has to be near zero for a low-energy state. We first consider the magnetic contribution 
\beq
\frac{1}{g^2} {\rm Tr} (F \wedge * F) , 
\eeq
where $F$ is the field strength (curvature) of the connection $A$ on the gauge bundle $P$. So in the weak coupling limit $g \rightarrow 0$, the wave-function of a low-energy state is concentrated near flat connections, for which $F = 0$. (See e.g. \cite{Witten00}.)

For a given $P$, there is a moduli space ${\cal M}$ of such flat connections. (For a general discussion, which however focuses on bundles over $T^3$, see \cite{Borel-Friedman-Morgan}.) In general, it consists of several components:
\beq
{\cal M} = \bigcup_\alpha {\cal M}_\alpha ,
\eeq
where the range of the label $\alpha$ depends on the topological class of $P$ as described by the discrete abelian magnetic 't~Hooft flux $m \in H^2 (T^4, C)$. (The instanton number $k \in H^4 (T^4, \mathbb Q)$ has to vanish for a flat connection, since its image in de~Rham cohomology is given by the class of ${\rm Tr} (F \wedge F)$.) Each component ${\cal M}_\alpha$ of ${\cal M}$ is of the form
\beq
{\cal M}_\alpha = (T^{r_\alpha} \times T^{r_\alpha} \times T^{r_\alpha} \times T^{r_\alpha}) / W_\alpha
\eeq
for some number $r_\alpha$, known as the rank of the component, and some discrete group $W_\alpha$, which acts on the torus $T^{r_\alpha}$. The simplest example of such a component is obtained for an arbitrary group $\Gadj$ by considering a topologically trivial bundle $P$, i.e. $0 = m \in H^2 (T^4, C)$. There is then a component ${\cal M}_0$ for which $r_0$ equals the rank of $\Gadj$. In fact, $T^{r_0}$ may be identified with a maximal torus of $\Gadj$, and $W_0$ is the corresponding Weyl group. But even for $m = 0$, there are in general also other components.

A flat connection is characterized by its holonomies
\beq
U \in {\rm Hom} (\pi_1 (T^4), G_{\rm adj}) ,
\eeq
modulo conjugation by elements of $G_{\rm adj}$ (which represent transformations in the connected component $\Omega_0$ of the group $\tilde{\Omega}$ of gauge transformations). After a choice of basis $\gamma_i$, $i = 1, \ldots 4$ of the homology group $H_1 (T^4, \Z) \simeq \Z^4$, we may represent $U$ by four commuting elements $U_i \in \Gadj$ modulo simultaneous conjugation by elements of $\Gadj$. An arbitrary lifting $\hat{U}_i \in G$ to the simply connected covering group is however only almost commuting, in the sense that
\beq \label{ACR}
\hat{U}_i \hat{U}_j \hat{U}_i^{-1} \hat{U}_j^{-1} = m_{ij} \in C .
\eeq
Here $m_{ij} \in H^2 (T^2, C)$ denotes the restriction of the discrete abelian magnetic 't~Hooft flux $m \in H^2 (T^4, C)$ to the two-torus spanned by the directions $i$ and $j$, i.e. $m = \frac{1}{2!} m_{ij} d x^i \wedge d x^j$, where the $d x^i$, constitute a basis of $H^1 (T^4, \Z)$ dual to the basis $\gamma_i$. Large gauge transformations act on the $\hat{U}_i$ by multiplication by elements of the center $C$, i.e. they amount to a change of lifting of the $U_i$ from $G_{\rm adj}$ to $G$.

At a point $p$ in a component ${\cal M}_a$ of rank $r_\alpha$ of the moduli space of flat connections, the holonomies $U_i$  spontaneously break the gauge symmetry to a subgroup of rank $r_\alpha$. Generically, the Lie algebra $h$ of this unbroken group is abelian, but in general it may be of the form
\beq
h \simeq s \oplus u(1)^r ,
\eeq
for some $r$, $0 \leq r \leq r_\alpha$, and some semi-simple algebra $s$ of rank $r_\alpha - r$.

Given such an algebra $h$, we let ${\cal M}^h \subset {\cal M}$ denote the closure of the corresponding subspace of ${\cal M}$. In general, ${\cal M}^h$ consists of several connected components:
\beq
{\cal M}^h = \bigcup_a {\cal M}^h_a ,
\eeq
where the range of the index $a$ depends on the subalgebra $h$ under consideration. Large gauge transformations obviously leave $h$ invariant, and thus act by permuting the components ${\cal M}_a$.

To begin with, we will consider the degrees of freedom associated with the semi-simple term $s$ in the unbroken symmetry algebra $h \simeq s \oplus u (1)^r$. The corresponding part of the holonomies parametrizes the directions in the moduli space ${\cal M}$ of flat connections that are normal to the submanifold ${\cal M}^h$ on which $h$ is restored. 

It is convenient to rescale the four spatial components $A_i$ of the gauge field and instead use the canonically normalized variables $A_i^\prime = g^{-1} A_i$. In the weak coupling limit $g \rightarrow 0$, the periodicity of the $A^\prime_i$ then goes to infinity, and they can be regarded as four ordinary scalar fields. Together with the five original scalar fields $\Phi^5, \ldots, \Phi^9$ and the fermionic fields $\Psi^1, \ldots, \Psi^4$, they constitute the degrees of freedom of supersymmetric quantum mechanics with 16 supercharges based on the Lie algebra $s$. This theory is the dimensional reduction (not compactification) of the corresponding Yang-Mills theory to $0+1$ dimensions.

This supersymmetric quantum mechanical model has no mass-gap, but is believed to have a finite dimensional linear space $V_s$ of normalizable zero energy states. In a perturbed version of the theory, known as $N = 1^*$ and obtained by adding a mass-term to $N = 4$ supersymmetric Yang-Mills theory in $d = 3 + 1$ dimensions, one finds that $V_s$ has an orthonormal basis with elements in one-to-one correspondence with the set of distinguished markings of the $s$ Dynkin diagram \cite{Kac-Smilga}: 

A marking of a Dynkin diagram defines a grading 
\beq
s = \bigoplus_{n \in \Z} s_n
\eeq
 of the Lie algebra, for which the simple roots corresponding to the marked and unmarked nodes have grading $+1$ and $0$ respectively, and the Cartan generators have grading $0$. The marking is called distinguished if 
\beq
 \dim s_0 = \dim s_{+1} = \dim s_{-1} . 
 \eeq
 Every Dynkin diagram admits a canonical distinguished marking in which all nodes are marked, so that $\dim s_0 = \dim s_{+1} = \dim s_{-1} = {\rm rank} \; s$, but  there are also other distinguished markings (see e.g. \cite{Collingwood-McGovern}):
\beq
 \dim V_s = \left\{
\begin{array}{ll}
1, & s \simeq su (n) \cr
\# \; {\rm partitions \; of} \; n \\
{\rm into \; distinct \; odd \; parts}, & s \simeq so (n) \cr
\# \; {\rm partitions \; of}  \; 2n \\
{\rm into \; distinct \; even \; parts}, & s \simeq sp (2 n) \cr
3, & s \simeq E_6 \cr
6, & s \simeq E_7 \cr
11, & s \simeq E_8 \cr
4, & s \simeq F_4 \cr
2, & s \simeq G_2 .
\end{array}
\right.
\eeq
A priori, one may think that adding a mass-perturbation would change $\dim V_s$, but these results are actually also (almost) uniquely determined by $S$-duality of $N = 4$ supersymmetric Yang-Mills theory on $T^3 \times \mathbb R$ with various (not necessarily simply laced) gauge groups \cite{Henningson-Wyllard}.

We must also consider the degrees of freedom associated with the abelian term $u (1)^r$ in the unbroken symmetry algebra $h \simeq s \oplus u (1)^r$. The corresponding part of the holonomies parametrizes the directions along the submanifold ${\cal M}^h$ on which $h$ is restored. 

Beginning with the gauge field $A$, the canonical conjugates to the holonomies are the components of the electric field strength $E_i = \frac{1}{g^2} \dot{A}_i$. These appear in the electric contribution 
\beq
g^2 {\rm Tr} (E_i E_i)
\eeq
to the Yang-Mills energy density, so a low-energy state must have $E_i = 0$. This means that the wave-function of such a state must be locally constant on ${\cal M}^h$, i.e. it must be constant on each component ${\cal M}^h_a$.

Continuing with the scalar fields $\Phi^5, \ldots, \Phi^9$, we need only consider the modes which are covariantly constant over $T^4$ with respect to the connection $A$, since non-constant modes necessarily carry energy of the order of the inverse size of the $T^4$. We denote the canonical conjugates of these constant modes as $\Pi^5, \ldots, \Pi^9$. The quantum theory will have a continuum of (non-normalizable unless $r = 0$) states $\left| \Pi_5, \ldots , \Pi_9 \right>$ labeled by the $5 r$ "eigenvalues" of the corresponding operators $\hat{\Pi}_5, \ldots, \hat{\Pi}_9$. We call this a rank $r$ continuum. The energy of these states is given by the term 
\beq
{\rm Tr} (\Pi^5 \Pi^5 + \ldots + \Pi^9 \Pi^9) 
\eeq
in the Yang-Mills energy density.

Finally we must take the spinor fields $\Psi^1, \ldots, \Psi^4$ into account. Again, we need only consider the covariantly constant modes, which however are their own canonical conjugates and generate a Clifford algebra. Quantization thus gives an additional $2^8$-fold degeneracy of the continua of states.

\subsection{Interpretation in $(2, 0)$ theory}
We have found that the complete low energy spectrum may be determined as follows: Each possible value of the discrete abelian magnetic 't~Hooft flux $m \in H^2 (T^4, C)$ determines a moduli space ${\cal M}$ of flat connections. Each possible unbroken subalgebra $h \simeq s \oplus u (1)^r$ is restored on a submanifold ${\cal M}^h$ of ${\cal M}$, every connected component ${\cal M}^h_a$ of which contributes a $\dim V_s$ dimensional vector space of rank $r$ continua of low-energy states (times the $2^8$-fold degeneracy due to the fermionic degrees of freedom). Large gauge transformations permute the components ${\cal M}^h_a$ while preserving $h$, and acts trivially on the set of distinguished markings of the $s$ Dynkin diagram. Diagonalizing this action on the total space of states gives a decomposition of the spectrum into states of definite discrete abelian electric 't~Hooft flux $e \in H^3 (T^4, C)$. The spectrum can be summarized by giving the number $N_{(m, e)}^r (\Gadj)$ of rank $r$ continua of discrete abelian magnetic and electric 't~Hooft fluxes $m$ and $e$ in the theory with gauge group $\Gadj$. 

We will now interpret the results as pertaining to type $\Phi$ $(2, 0)$ theory on $T^5 \times \mathbb R$, where  $T^5 = T^4 \times S^1$: The procedure described so far is covariant with respect to the $\SL_4 (\Z)$ mapping class group of $T^4$, so the numbers $N_{(m, e)}^r (\Gadj)$ manifestly depend on $m$ and $e$ only via the $\SL_4 (\Z)$ orbit of the pair $(m, e)$. But by the K\"unneth isomorphism
\beq
H^3 (T^5, C) \simeq H^2 (T^4, C) \oplus H^3 (T^4, C) ,
\eeq
the discrete abelian magnetic and electric 't~Hooft fluxes $m \in H^2 (T^4, C)$ and $e \in H^3 (T^4, C)$ can be seen as components of a single characteristic class 
\beq
f = m + e \in H^3 (T^5, C) ,
\eeq
that we will simply call the discrete abelian 't~Hooft flux. We thus write
\beq
N_f^r (\Phi) = N_{(m, e)}^r (\Gadj) ,
\eeq
and $(2, 0)$ theory predicts that these numbers should actually only depend on $f \in H^3 (T^5, C)$ via its $\SL_5 (\Z)$ orbit $[f]$. In general, this prediction is non-trivial, since a single $\SL_5 (\Z)$ orbit $[f]$ may consist of several $\SL_4 (\Z)$ orbits of the pair $(m, e)$.

In the following sections, we will verify this prediction in detail for the $A$- and $D$-series, leaving the three $E$-type models for the future. There is, however, one check that can be performed without having to specify precisely which $(2, 0)$ theory $\Phi$ we are considering. To describe this, we begin by recalling that low-energy states are localized on flat connections, for which the instanton number $k = \left[{\rm Tr} (F \wedge F) \right]$ vanishes, and thus in particular is integer-valued. In view of the relation (\ref{fraction}), this means that
\beq \label{mm-condition}
0 = \frac{1}{2} m \cdot m \in H^0 (T^4, \mathbb Q) \mod H^0 (T^4, \Z) 
\eeq
for a low-energy state. Next, we fix $m \in H^2 (T^4, C)$ and ask what values of $e \in H^3 (T^4, C)$ are possible for a low-energy state. As described above, $e$ determines the transformation properties of a state under the group $\Omega$ of large gauge transformations, which act on the holonomies by multiplication by a quartet of center elements $c_i \in C$. But if such a transformation is equivalent to simultaneous conjugation of the holonomies by some element $g \in G$ (i.e. a gauge transformation in the connected component $\Omega_0$), it is trivial and must thus be trivially represented. In particular, we may choose the element $g$ as one of the holonomies $U_i$, conjugation by which, in view of the almost commutation relations (\ref{ACR}), is equivalent to multiplication of the holonomies by certain quartets of center elements. The requirement that these transformations are trivially represented on the states gives a set of restrictions on the possible values of $e$, that can be summarized as
\beq \label{me-condition}
0 = m \cdot e \in H^1 (T^4, \mathbb Q) \mod H^1 (T^4, \Z)
\eeq
for a low-energy state. The necessary conditions (\ref{mm-condition}) and (\ref{me-condition}) for a low-energy state may now be summarized as an $\SL_5 (\Z)$ covariant condition on the characteristic class $f = m + e$:
\beq \label{ff-condition}
0 = \frac{1}{2} f \cdot f \in H^1 (T^5, \mathbb Q) \mod H^1 (T^5, \Z) .
\eeq

We may write out the formulas (\ref{mm-condition}, \ref{me-condition}, and \ref{ff-condition}) more explicitly by expanding $m$, $e$, and $f$ as
\bea
m & = & \frac{1}{2} m_{ij} d x^i \wedge d x^j \cr
e & = & \frac{1}{6} e_{ijk} d x^i \wedge d x^j \wedge d x^k \cr
f & = & \frac{1}{6} f_{abc} d x^a \wedge d x^b \wedge d x^c ,
\eea
where $i, j, k = 1, \ldots, 4$ and $a, b, c = 1, \ldots, 5$. We then have
\bea
\frac{1}{2} m \cdot m & = & \frac{1}{8} m_{ij} m_{kl} \epsilon^{ijkl} \cr
m \cdot e & = & \frac{1}{6} m_{ij} e_{klm} \epsilon^{jklm} d x^i \cr
f \cdot f & = & \frac{1}{24} f_{abc} f_{def} \epsilon^{bcdef} d x^a ,
\eea
where $\epsilon^{ijkl}$ and $\epsilon^{abcde}$ are totally anti-symmetric with  $\epsilon^{1234} = \epsilon^{12345} = 1$.

So (\ref{ff-condition}) is a necessary condition for low-energy states. In the following sections, we will explicitly compute the spectrum of the $A$- and $D$-series for these values of $f \in H^3 (T^5, C)$.

\setcounter{equation}{0}
\section{The $A$-series}
The simply connected group corresponding to the $\Phi = A_{n - 1}$ model is 
\beq
G = SU (n)
\eeq
 consisting of unitary unimodular $n \times n$ matrices. Its center subgroup $C \simeq \Z_n$ consists of matrices of the form $\exp (2 \pi i c / n) \id_n$ for $c \in \Z_n$. 

The $\SL_4 (\Z)$ orbit $[m]$ of $m = \frac{1}{2} m_{ij} d x^i \wedge d x^j \in H^2 (T^4, \Z_n)$ is completely classified by the invariants $u$ and $k^\prime$ defined by
\bea
u & = & {\rm gcd} (m_{ij}, n) \cr
k^\prime & = & \frac{1}{u^2} {\rm Pf} (m) .
\eea
Here ${\rm Pf} (m) \in \Z_n$ is the Pfaffian of $m$ defined by
\beq
 {\rm Pf} (m) = m_{12} m_{34} + m_{13} m_{42} + m_{14} m_{23}  .
\eeq
For given values of these invariants, a representative of the orbit $[m]$ is given by
\beq
m_{ij} = u \left(
\begin{array}{cccc}
0 & 1 & 0 & 0 \cr
-1 & 0 & 0 & 0 \cr
0 & 0 & 0 & k^\prime \cr
0 & 0 & -k^\prime & 0
\end{array}
\right) .
\eeq
We also define the integers $v$ and $v^\prime$ by
\bea
v & = & n / u \cr
v^\prime & = & v / {\rm gcd} (k^\prime, v) .
 \eea
  In terms of these variables, the condition (\ref{mm-condition}), that is necessary for a non-empty low-energy spectrum, takes the form $u^2 k^\prime = 0 \mod n$,  which is equivalent to demanding that the a priori rational number $w$ defined as
\beq
w = u / v^\prime
\eeq
be an integer.

For a given value of $w$, possible unbroken subalgebras $h$ of $su (n)$ are of the form
\beq
h = s \oplus u(1)^r \simeq su (w_1) \oplus \ldots \oplus su (w_{r+1}) \oplus u (1)^r ,
\eeq
for some partition
\beq
w_1 + \ldots w_{r + 1} = w  
\eeq
of $w$ into $r + 1$ parts. To describe the corresponding holonomies, we define $w^\prime$, $t$, and $t_1, \ldots, t_{r+1}$ by
\bea
w^\prime & = & {\rm gcd} (w_1, \ldots, w_{r+1}) \cr
t & = & w / w^\prime \cr
t_1 & = & w_1 / w^\prime \cr
& \ldots & \cr
t_{r+1} & = & w_{r+1} / w^\prime ,
\eea
so that
\beq
t_1 + \ldots + t_{r + 1} = t
\eeq
is a partition of $t$ into relatively prime parts. The holonomies may then be conjugated to a subgroup
\beq
SU (v) \otimes SU (v^\prime) \otimes SU (w^\prime) \otimes U (t) \subset SU (n) ,
\eeq
where they take the form 
\bea \label{A-holonomies}
U_1 & = & A \otimes \id_{v^\prime} \otimes \id_{w^\prime} \otimes T_1\cr
U_2 & = & B \otimes \id_{v^\prime} \otimes \id_{w^\prime} \otimes T_2 \cr
U_3 & = & \id_v \otimes A^\prime \otimes \id_{w^\prime} \otimes T_3 \cr
U_4 & = & \id_v \otimes B^\prime \otimes \id_{w^\prime} \otimes T_4 .
\eea
Here $A$ and $B$ are some fixed $SU (v)$ matrices that fulfill the almost commutation relations
\beq
A B A^{-1} B^{-1} = \exp (2 \pi i / v) \id_v ,
\eeq
and $A^\prime$ and $B^\prime$ are fixed $SU (v^\prime)$ matrices that obey
\beq
A^\prime B^\prime A^{\prime -1} B^{\prime -1} = \exp (2 \pi i k^\prime / v)  \id_{v^\prime} .
\eeq
We may for example choose
\beq
A = {\left( \begin{array}{cccc}
e^{i \pi (- v + 1) / v} & 0 & \ldots & 0 \\
0 & e^{i \pi (-v + 3) / v} & \ldots & 0 \\
\vdots & & \ddots & \vdots \\
0 & \ldots & \ldots &  e^{i \pi (v - 1) / v}
\end{array} \right) } 
\eeq
and
\beq 
B  =  {\left( \begin{array}{cccc} 
0 & 1 & \cdots & 0 \\
\vdots & 0 &\ddots & \vdots \\
0 & & 0 & 1 \\
(-1)^{v-1} & 0& \hdots & 0
\end{array} \right) }
\eeq
and similarly for $A^\prime$ and $B^\prime$. The $T_i$ are block-diagonal matrices of the form
\beq
T_i = {\rm diag} \left(\exp (i \phi_i^{(1)}) \id_{t_1}, \ldots, \exp (i \phi_i^{(r+1)}) \id_{t_{r+1}} \right) ,
\eeq
where the angular variables $\phi_i^{(1)}, \ldots, \phi_i^{(r+1)}$ are subject to the restriction that the complete matrices $U_i$ in (\ref{A-holonomies}) should be elements of $SU (n)$ (rather than of $U (n)$).

A large gauge transformation parametrized by $c = c_i d x^i \in H^1 (T^4, \Z_n)$ acts on the holonomies according to
\beq
U_i \mapsto e^{2 \pi i c_i / n} U_i .
\eeq
However, multiplication of $U_1$ or $U_2$ by $e^{2 \pi i / v}$ is equivalent to conjugation by the matrices $B$ or $A$ respectively, and multiplication of $U_3$ or $U_4$ by $e^{2 \pi i k^\prime / v}$ is equivalent to conjugation by $B^\prime$ or $A^\prime$. Furthermore multiplication of any $U_i$ by $e^{2 \pi i / t}$ can be absorbed by shifting the angles in the expression for $T_i$. This implies that transformations for which 
\bea
\frac{c_1}{n} & = & 0 \mod \frac{{\rm gcd} (v, t)}{v t} \cr
\frac{c_2}{n} & = & 0 \mod \frac{{\rm gcd} (v, t)}{v t} \cr
\frac{c_3}{n} & = & 0 \mod \frac{{\rm gcd} (v, k^\prime t)}{v t} \cr
\frac{c_4}{n} & = & 0 \mod \frac{{\rm gcd} (v, k^\prime t)}{v t} 
\eea
are trivial.

Since the number of normalisable states in $s \simeq su (w_1) \oplus \ldots \oplus su (w_{r+1})$ quantum mechanics with 16 supercharges is $\dim V_s = 1$, there is a single rank $r$ continuum of quantum states associated with each connected component ${\cal M}^h_a$ of the space of holonomies ${\cal M}^h$ with unbroken algebra $h$. These components are permuted by large gauge transformations. Diagonalizing this action, we find that there is one continuum for each value of the electric 't~Hooft flux $e = \frac{1}{6} e_{ijk} dx^i d x^j dx^k \in H^3 (T^4, \Z_n)$ such that the components satisfy the conditions
\bea
e_{234} & \in & \frac{v t}{{\rm gcd} (v, t)} \Z_n \cr
e_{134} & \in & \frac{v t}{{\rm gcd} (v, t)} \Z_n \cr
e_{124} & \in & \frac{v t}{{\rm gcd} (v, k^\prime t)} \Z_n \cr
e_{123} & \in & \frac{v t}{{\rm gcd} (v, k^\prime t)} \Z_n .
\eea

We may now describe the low-energy spectrum in an $\SL_5 (\Z)$ covariant form as appropriate for the $(2, 0)$ theory of type $\Phi = A_{n - 1}$: Let
\beq
n_1 + \ldots + n_{r + 1} = n
\eeq
be a partition of $n$ into $r + 1$ parts.  To these data is associated a set of rank $r$ continua of states, one for every value of $f = m + e \in H^3 (T^5, \Z_n)$ that is divisible by 
\beq
t = \frac{n}{{\rm gcd} (n_1, \ldots, n_{r+1})}
\eeq
and obeys the constraint (\ref{ff-condition}).

\setcounter{equation}{0}
\section{The $D$-series}
The simply connected group corresponding to type $\Phi = D_n$ is 
\beq
G = \Spin (2 n) , 
\eeq
i.e. the universal double cover of the group $\SO (2 n)$ of unimodular orthogonal $2 n \times 2 n$ matrices. The center $C$ of $G$ is
\beq
C = \{ 0, v, s, c \} .
\eeq
Here the trivial element $0$ and the non-trivial element $v$ (for "vector") project to the identity element $\id_{2 n}$ of $\SO (2 n)$, whereas the non-trivial elements $s$ and $c$ (for "spinor" and "cospinor" respectively) project to the non-trivial center element $-\id_{2 n}$ of $\SO (2 n)$. Our first aim is to describe the $\SL_4$ orbits $[m]$ of $m \in H^2 (T^4, C)$ and how they give rise to $\SL_5 (\Z)$ orbits $[f]$ of $f = m + e \in H^3 (T^5, C)$ by choosing different values of $e \in H^3 (T^4, C)$. 

To begin with, we consider the quotient of $C$ by the subgroup generated by $v$. This quotient group is isomorphic to $\Z_2$. We let $m^v \in H^2 (T^4, \Z_2)$ denote the corresponding reduction of $m \in H^2 (T^4, C)$ modulo $v$:
\beq
m^v = m \mod v .
\eeq
There are $3$ different $\SL_4 (\Z)$ orbits of $m^v$, that we denote as $[m^v] = 0, 1, 2$ respectively. These symbols $0, 1, 2$ are fairly arbitrary, but could be interpreted as the number of obstructions to having a vector structure on the bundle. We say that the bundles with $[m^v] = 0, 1, 2$ have vector structure, half vector structure, and no vector structure respectively. The following table gives the equations defining these orbits and their cardinalities:
\beq
\begin{array}{rlr}
& & \# \cr
\hline
[m^v]  = 0 & m^v = 0 & 1 \cr
1 & m^v \neq 0, {\rm Pf} (m^v) = 0 &  35 \cr
2 &  {\rm Pf} (m^v) = 1 & 28 \cr
\hline
& & 64
\end{array}
\eeq

For the bundles with vector structure, i.e. the $[m^v] = 0$ bundles, we have 
\beq
m = v \, \tilde{m}
\eeq
for some $\tilde{m} \in H^2 (T^2, \Z_2)$. The $\SL_4 (\Z)$ orbit of $m$ is thus determined by the orbit $[\tilde{m}]$ of $\tilde{m}$, which we denote with the symbols $0, 1, 2$ introduced in the previous paragraph. 

For the bundles with half vector structure, i.e. the $[m^v] = 1$ bundles, and the bundles with no vector structure, i.e. the $[m^v] =  2$ bundles, we have to treat the cases of $n$ odd and $n$ even separately.

\subsection{$n$ odd}
When $n$ is odd, $C \simeq \Z_4$ with the following identifications:
\bea
0 & = & 0 \cr
s & = & 1 \cr
v & = & 2 \cr
c & = & 3 .
\eea
For $m \in H^2 (T^4, \Z_4)$, we thus have
\beq
m^v = m \mod 2 \in H^2 (T^4, \Z_2) .
\eeq
There are $7$ different $\SL_4 (\Z)$ orbits of $m \in H^2 (T^4, \Z_4)$ that we denote as follows: 
\bea
[m] & = & 0, 0^\prime, 0^{\prime \prime} , \cr
& & 1, 1^\prime \cr
& & 2, 2^\prime .
\eea
The orbits $[m]Ê= 0, 0^\prime, 0^{\prime \prime}$ have vector structure, so that $m = v \, \tilde{m}$ for some $\tilde{m} \in H^2 (T^4, \Z_2)$. They have $[\tilde{m}] = 0, 1, 2$ respectively. The orbits $[m] = 1, 1^\prime$ both have half vector structure, i.e. $[m^v] = 1$, and are distinguished by having $\Pf (m) = 0, 2$ modulo $4$ respectively. The orbits $[m] = 2, 2^\prime$ both have no vector structure, i.e. $[m^v] = 2$, and are distinguished by having $\Pf (m) = 1, 3$ modulo $4$ respectively. 

Choosing an arbitrary representative $m \in H^2 (T^2, \Z_4)$ of an $\SL_4 (\Z)$ orbit $[m]$ and taking $e = 0$ in $H^3 (T^4, \Z_4)$ gives rise to an $f = m + e \in H^3 (T^5, \Z_4)$, whose $\SL_5 (\Z)$ orbit $[f]$ we denote with the same symbol as the orbit $[m]$. However, the two orbits $[m] = 2$ and $[m] = 2^\prime$ give rise to a single orbit, which we denote as $[f] = 2$. Taking a non-vanishing value of $e$ may give rise to another orbit $[f]$, i.e. not the orbit denoted by the same symbol as the orbit $[m]$. But it turns out that no further $\SL_5 (\Z)$ orbits appear, i.e. the complete list is
\bea
[f] & = & 0, 0^\prime, 0^{\prime \prime}, \cr
& & 1, 1^\prime, \cr
& & 2 .
\eea
We say that the $[f]$ in the first, second, and third line have vector structure, half vector structure, and no vector structure respectively. In the following table, the columns correspond to the $\SL_4 (\Z)$ orbits $[m]$, and the rows correspond to the $\SL_5 (\Z)$ orbits $[f]$. The symbols and the cardinalities of the orbits are given in the first and last row or column respectively. The entries are the number of values of $e \in H^3 (T^4, \Z_4)$ that give rise to a $f \in H^3 (T^5, \Z_4)$ in a certain orbit $[f]$ for a given representative $m \in H^2 (T^4, \Z_4)$ of an orbit $[m]$:
{\small
\beq
\begin{array}{rrrrrrrrr}
& [m] = 0 & 0^\prime  & 0^{\prime \prime}  & 1 & 1^\prime & 2&  2^\prime & \# \cr
\hline
[f] = 0 & 1 & & & & & & & 1 \cr
0^\prime & 15 & 4 & & & & & & 155 \cr
0^{\prime \prime}  & & 12 & 16 & & & & & 868 \cr
1 & 240 & 48 & & 16 & & & & 19840 \cr
1^\prime & & 192 & 240 & 48 & 64 & & & 138880 \cr
2 & & & & 192 & 192 & 256 & 256 & 888832 \cr
\hline
\# & 1 & 35 & 28 & 1120 & 1120 & 896 & 896 &
\end{array}
\eeq
}
The inner product on $C$ is determined by 
\beq
s \cdot s = \frac{1}{4} \mod 1 .
\eeq
It follows that an $f \in H^3 (T^5, C)$ such that
\beq
[f] = 1^\prime, 2, 
\eeq
has a non-vanishing value of $f \cdot f \in H^1 (T^5, C)$, and thus an empty low-energy spectrum.

\subsection{$n$ even}
When $n$ is even, $C \simeq \Z_2 \times \Z_2$ with the following identifications:
\bea
0 & = & (0,0) \cr
s & = & (1,0) \cr
v & = & (1,1) \cr
c & = & (0,1) .
\eea
The theory is invariant under an $S_2$ group of outer automorphism which permutes the elements $s$ and $c$ while leaving $v$ invariant.

We decompose $m \in H^2 (T^4, Z_2 \times \Z_2)$ uniquely as 
\beq
m = s \, m^s + c \, m^c 
\eeq
with $m^s, m^c \in H^2 (T^2, \Z_2)$. We then have 
\beq
m^v = m^s + m^c \in H^2 (T^4, \Z_2) . 
\eeq
We denote the $\SL_4 (\Z)$ orbits of $m^s$, $m^c$, $m^v \in H^2 (T^4, \Z_2)$ by $[m^s], [m^c], [m^v] = 0, 1, 2$ as described above. The $\SL_4 (\Z)$ orbit of $m$ is completely determined by these, and we denote it as
\beq
[m] = [m^s] [m^c] [m^v] .
\eeq
Not all combinations of $[m^s]$, $[m^c]$, and $[m^v]$ are possible, though, and the complete list is
\bea
[m] & = & 000,110,220, \cr
& & 011,101,111,221,211,121, \cr
& & 022,202,112,122,212,222 ,
\eea
with the $[m]$ in the first, second, and third lines having vector structure, half vector structure, and no vector structure respectively. For the cases with vector structure, $\tilde{m} \in H^2 (T^4, \Z_2)$ is given by
\beq
\tilde{m} = m^s = m^c .
\eeq

Again, taking $e = 0$ in $H^3 (T^4, \Z_2 \times \Z_2)$ for some representative $m \in H^2 (T^4, Z_2 \times \Z_2)$ of an $\SL_4 (\Z)$ orbit $[m]$ gives an $f = m + e \in H^3 (T^5, Z_2 \times \Z_2)$ whose $\SL_5 (\Z)$ orbit $[f]$ we denote with the same symbol as $[m]$. All orbits $[f]$ obtained in this way are distinct. Taking a non-vanishing value of $e$ may give rise to another orbit $[f]$, i.e. not the orbit denoted by the same symbol as $[m]$. In this way, some further orbits appear: The orbit $[m] = 222$ gives rise not only to the orbit $[f] = 222$, but also to another orbit which we denote as $[f] = 222^\prime$. Similarly, each of the orbits $[m]Ê= 122, 212, 221$ give rise not only to $[f]Ê= 122,  212, 221$ respectively, but also to $[f] = 222^\prime$ and to a further orbit that we denote as $[f]Ê= 122, 212^\prime, 221^\prime$ respectively. The complete list of $\SL_5 (\Z)$ orbits is thus
\bea
[f] & = & 000,110,220, \cr
& & 011,101,111,221,211,121, \cr
& & 022, 202,112, 122,212, 222, 122^\prime, 212^\prime, 221^\prime, 222^\prime .
\eea
We say that $[f]$ in the first, second, and third row has vector structure, half vector structure, and no vector structure respectively. In the following table, the columns correspond to the $\SL_4 (\Z)$ orbits $[m]$, and the rows correspond to the $\SL_5 (\Z)$ orbits $[f]$. The symbols and the cardinalities of the orbits are given in the first and last row or column respectively. The entries are the number of values of $e \in H^3 (T^4, \Z_2 \times \Z_2)$ that give rise to an $f = m + e \in H^3 (T^5, \Z_2 \times \Z_2)$ in a certain orbit $[f]$ for a given representative $m \in H^2 (T^4, \Z_2 \times \Z_2)$ of an orbit $[m]$.
\beq
{\tiny
\begin{array}{rrrrrrrrrrrrrrrrrrrrrrrrr}
& [m] = 000 & 110 & 220 & 011 & 101 & 111 & 221 & 211  & 121 & 022 & 202 & 112 & 122 & 212 & 222 & \# \cr
\hline
[f] = 000 & 1 & & & & & & & & & & & & & & & 1 \cr
110 & 15 & 4 & & & & & & & & & & & & & & 155 \cr
220 & & 12 & 16 & & & & & & & & & & & & & 868 \cr
011 & 15 & & & 4 & & & & & & & & & & & & 155 \cr
101 & 15 & & & & 4 & & & & & & & & & & & 155 \cr
111 & 210 & 12 & & 12 & 12 & 8 & & & & & & & & & & 6510 \cr
221 & & 36 & & & & 8 & 16 & & & & & & & & & 13020 \cr
211 & & 48 & & & 48 & 8 & & 16 & & & & & & & & 17360 \cr
121 & & 48 & & 48 & & 8 & & & 16 & & & & & & & 17360 \cr
022 & & & & 12 & & & & & & 16 & & & & & & 868 \cr
202 & & & & & 12 & & & & & & 16 & & & & & 868 \cr
112 & & & & 48 & 48 & 8 & & & & & & 16 & & & & 17360 \cr
122 & & & & 36 & & 8 & & & & & & & 16 & & & 13020 \cr
212 & & & & & 36 & 8 & & & & & & & & 16 & & 13020 \cr
222 & & & & & & 8 & & & & & & & & & 16 & 10416 \cr
122^\prime & & & & 96 & & 32 & & & 48 & 240 & & 48 & 48 & & & 104160 \cr
212^\prime & & & & & 96 & 32 & & 48 & & & 240 & 48 & & 48 & & 104160 \cr
221^\prime & & 96 & 240 & & & 32 & 48 & 48 & 48 & & & & & & & 104160 \cr
222^\prime & & & & & & 96 & 192 & 144 & 144 & & & 144 & 192 & 192 & 240 & 624960 \cr
\hline
\# & 1 & 35 & 28 & 35 & 35 & 630 & 420 & 560 & 560 & 28 & 28 & 560 & 420 & 420 & 336 &
\end{array}
}
\eeq
The inner product on $C$ depends on whether $n = 2$ or $n = 0$ modulo $4$: In the first case,  we have
\beq
\begin{array}{ccc}
& s & c \cr
\hline
s & \frac{1}{2} & 0 \cr
c & 0 & \frac{1}{2} 
\end{array}
\mod 1,
\eeq
and in the second case,
\beq
\begin{array}{ccc}
& s & c \cr
\hline
s & 0 & \frac{1}{2} \cr
c & \frac{1}{2} & 0 
\end{array}
\mod 1 .
\eeq
It follows that the orbits
\bea
[f] & = & 211, 121, \cr
& & 112, 222, 122^\prime, 212^\prime, 221^\prime, 222^\prime
\eea
for $n = 0 \mod 4$, and the orbits
\bea
[f] & = & 211, 121, \cr
& & 022, 202, 122, 212, 122^\prime, 212^\prime, 221^\prime, 222^\prime
\eea
for $n = 2 \mod 4$, have empty low-energy spectra. The orbits
\bea
[f] & = & 211, 121, \cr
& & 122^\prime, 212^\prime, 221^\prime, 222^\prime 
\eea
thus have empty low-energy spectra for all even $n$. 

\subsection{The generating functions}
Our aim is to compute the degeneracies $N_f^r (D_n)$ of rank $r$ continua with characteristic class $f \in H^3 (T^5, C)$ in the $\Phi = D_n$ theory. It is convenient, however, to treat all values of $n$ and $r$ simultaneously by introducing a set of generating functions $Z_f$ defined as
\beq
Z_f (q, y) = \sum_{k = 0}^\infty \sum_{r = 0}^\infty N_f^r (D_{2 k + 1}) q^{4 k + 2} y^r 
\eeq
for the case of $n$ odd, and
\beq
Z_f (q, y) = \sum_{k = 0}^\infty \sum_{r = 0}^\infty N_f^r (D_{2 k}) q^{4 k} y^r 
\eeq
for the case of $n$ even. We will compute these functions in terms of some other functions $P$, $Q$, and $R$, that we will now define. 

$P$ and $Q$ are the generating functions for the dimensions of the spaces $V_s$ of normalizable zero-energy states in $s$ quantum mechanics with $16$ supercharges for $s$ an orthogonal or symplectic algebra:
\bea 
P (q) & = & \sum_{n = 0}^\infty \dim V_{\so (n)} q^n = \prod_{k = 1}^\infty (1 + q^{2 k -1}) \cr
& = & 1+ q + q^3 + q^4 + q^5 + q^6 + q^7 + 2 q^8 + \ldots 
\eea
and
\bea 
Q (q) & = & \sum_{n = 0}^\infty \dim V_{\symp (2n)} q^{2 n} = \prod_{k = 1}^\infty (1 + q^{2 k}) \cr
& = & 1 + q^2 + q^4 + 2 q^6 + 2 q^8 + 3 q^{10} + \ldots
\eea
It will be convenient to decompose $P (q)$ into terms $\Pe (q)$ and $\Po (q)$ with even and odd powers of $q$ respectively, i.e.
\bea \label{Pe-Po}
\Pe (q) & = & \frac{1}{2} \left( P (q) + P (-q) \right) \cr
\Po (q) & = & \frac{1}{2} \left( P (q) + P (-q) \right) .
\eea
$R$ is a generating function for the number $N_k^r$ of conjugacy classes of subalgebras of $\su (k)$ with abelian term $u (1)^{r-1}$. This is given by the number of partitions of $k$ into $r$ parts, so we define
\bea \label{R}
R (q, y) & = & \sum_{k = 1}^\infty \sum_{r = 1}^\infty N_k^r y^r q^{2 k} = \prod_{k = 1}^\infty (1 - y q^{2 k})^{-1} \cr
& = & 1 + y q^2 + (y + y^2) q^4 + (y + y^2 + y^3) q^6 + (y + 2 y^2 + y^3 + y^4) q^8 + \ldots \cr
& & 
\eea

\subsection{Bundles with vector structure}
As described above, these bundles have $m^v = m \mod v \in H^2 (T^4, \Z_2)$ vanishing, i.e. $[m^v] = 0$. Thus $m = v \tilde{m}$ for some $\tilde{m} \in H^2 (T^4, \Z_2)$, and the orbits are determined by $[\tilde{m}] = 0, 1, 2$. This corresponds to
\beq
[m] = 0, 0^\prime, 0^{\prime \prime}
\eeq
for $n$ odd, and 
\beq
[m] = 000, 110, 220 
\eeq
for $n$ even. 

The $SO (2 n)$ projections of the holonomies $U_i$ may be conjugated to a subgroup
\beq
\SO (2 s) \times \SO (2 n - 2 s) \subset \SO (2 n) ,
\eeq
for certain values of $s$, $0 \leq s \leq n$, and then take the form 
\beq
U_i = (V_i, T_i)  .
\eeq
Here the $V_i$ are given by
\bea \label{Vcolumns}
V_1 & = & {\rm diag} (+\id,+\id,+\id,+\id,+\id,+\id,+\id,+\id,-\id,-\id,-\id,-\id,-\id,-\id,-\id,-\id) \cr
V_2 & = & {\rm diag} (+\id,+\id,+\id,+\id,-\id,-\id,-\id,-\id,+\id,+\id,+\id,+\id,-\id,-\id,-\id,-\id) \cr
V_3 & = & {\rm diag} (+\id,+\id,-\id,-\id,+\id,+\id,-\id,-\id,+\id,+\id,-\id,-\id,+\id,+\id,-\id,-\id) \cr
V_4 & = & {\rm diag} (+\id,-\id,+\id,-\id,+\id,-\id,+\id,-\id,+\id,-\id,+\id,-\id,+\id,-\id,+\id,-\id) \cr
& & 
\eea
with some multiplicities 
\beq
k, k_4, k_3, k_{34}, k_2, k_{24}, k_{23}, k_{234}, k_1, k_{14}, k_{13}, k_{134}, k_{12}, k_{124}, k_{123}, k_{1234}
\eeq
of the $16$ columns. It is convenient to think of these columns as associated with the points of the four-dimensional vector space $\Z_2^4$ over the finite field $\Z_2$. Like many of the constructions in the rest of the paper, this may be interpreted in terms of orientifolds. (See e.g. \cite{Witten97} \cite{Keurentjes}.)

The $T_i$ can be chosen in a $(\SO (2))^{n - s}$ subgroup, i.e. a maximal torus, of $\SO (2 n - 2 s)$. In each factor, they are parametrized by a quartet  $(\theta_1, \theta_2, \theta_3, \theta_4)$ of angular variables and take the form
\beq
T_i = \left( \begin{array}{cc}
\cos \theta_i & \sin \theta_i \cr
- \sin \theta_i & \cos \theta_i 
\end{array} \right) .
\eeq
Generically, the unbroken subalgebra is $h \simeq u(1)^{n - s}$. However, if $l$ of the quartets $(\theta_1, \theta_2, \theta_3, \theta_4)$ of angular variables are equal, a factor $u (1)^l$ gets enhanced to ${\rm u} (l) \simeq \su (l) \oplus u (1)$. If furthermore this common value of the $l$ quartets is such that all entries are equal to $0$ or $\pi$, the corresponding factors of the $T_i$ equal one of the $16$ special values in (\ref{Vcolumns}) with multiplicity $2 l$. There is then further enhancement to $\so (2 l + k^\prime)$, where $k^\prime$ is the multiplicity associated with that point in $\Z_2^4$. So only the modulo $2$ reductions of the multiplicities are invariant under continuous deformations of the angles $\theta_i$.

The modulo $2$  reductions of the $16$ multiplicities are constrained by the following $11$ linear equations over $\Z_2$:
\bea \label{n-equations}
0 & = & k {+} k_4 {+} k_3 {+} k_{34} {+} k_2 {+} k_{24} {+} k_{23} {+} k_{234} \cr
& & {+} k_1 {+} k_{14} {+} k_{13} {+} k_{134} {+} k_{12} {+} k_{124} {+} k_{123} {+} k_{1234} \cr
0 & = & k_1 {+} k_{14} {+} k_{13} {+} k_{134} {+} k_{12} {+} k_{124} {+} k_{123} {+} k_{1234} \cr
0 & = & k_2 {+} k_{24} {+} k_{23} {+} k_{234} {+} k_{12} {+} k_{124} {+} k_{123} {+} k_{1234} \cr
0 & = & k_3 {+} k_{34} {+} k_{23} {+} k_{234} {+} k_{13} {+} k_{134} {+} k_{123} {+} k_{1234} \cr
0 & = & k_4 {+} k_{34} {+} k_{24} {+} k_{234} {+} k_{14} {+} k_{134} {+} k_{124} {+} k_{1234} \cr
\tilde{m}_{12} & = & k_{12} {+} k_{124} {+} k_{123} {+} k_{1234} \cr
\tilde{m}_{13} & = & k_{13} {+} k_{134} {+} k_{123} {+} k_{1234} \cr
\tilde{m}_{14} & = & k_{14} {+} k_{134} {+} k_{124} {+} k_{1234} \cr
\tilde{m}_{23} & = & k_{23} {+} k_{234} {+} k_{123} {+} k_{1234} \cr
\tilde{m}_{24} & = & k_{24} {+} k_{234} {+} k_{124} {+} k_{1234} \cr
\tilde{m}_{34} & = & k_{34} {+} k_{234} {+} k_{134} {+} k_{1234} .
\eea
The first equation expresses the fact that the sum $2 s$ of the multiplicities is even. The next four equations follow from the requirement that the $V_i$ should be elements of $\SO (2 s)$ (rather than ${\rm O} (2 s)$). The last six equations ensure that the holonomies obey the correct almost commutation relations. There are $2^{16-11} = 32$ solutions to these equations, each of which is characterized by the number $2 d$, $0 \leq 2 d \leq 16$, of multiplicities that are equal to $1$ modulo $2$. We get the following number of solutions, depending on the values of $[\tilde{m}]$ and $2 d$:
\beq
\begin{array}{rrrrrrrrrr}
& 2 d = 0 & 2 & 4 & 6 & 8 & 10 & 12 & 14 & 16 \cr
\hline
[\tilde{m}] = 0 & 1 & & & & 30 & & & & 1 \cr
1 & & & 4 & & 24 & & 4 & & \cr
2 & & & & 16 & & 16 & & & \cr
\end{array}
\eeq
Recalling that $\dim V_{\su (l)} = 1$ and that $\dim V_{\so (n)}$ for $n$ even and odd are described by the generating functions $\Pe$ and $\Po$ respectively introduced in (\ref{Pe-Po}), we see that the generating function $Z$ for the total number of continua of states associated with such a solution is given by
\beq
R (q, y) \Po^{2 d} (q) \Pe^{16 - 2 d} (q) ,
\eeq
where the function $R$ is defined in (\ref{R}). Here, the coefficient of $y^r q^{2 n}$ is given by the number of rank $r$ continua in the $\Phi = D_n$ theory. 

\subsubsection{Electric 't~Hooft flux with vector structure}
Generically, each solution to (\ref{n-equations}) corresponds to a single component ${\cal M}_\alpha$ of rank $r_\alpha = n - d$ of the moduli space ${\cal M}$ of flat connections. This is true if the different liftings of the holonomies from $\SO (2 n)$ to $\Spin (2 n)$, i.e. large gauge transformations parametrized by $v \in C$, are equivalent modulo simultaneous conjugation of the holonomies, and implies that $e \in H^3 (T^4, C)$ will be of the form 
\beq
e = v \, \tilde{e}
\eeq
 for some $\tilde{e} \in H^3 (T^4, \Z_2)$. We say that such an $e$ has vector structure. 

Large gauge transformations parametrized by $s \in C$ (or equivalently by $c \in C$, since we still assume that $v = s - c$ acts trivially) act by interchanging the points of $\Z_2^4$ pairwise. There are $15$ non-trivial such transformations, and generically they act freely, but it might be that a particular transformation has a fixed point. A necessary condition for this to happen is that the multiplicities are pairwise equal. For example, symmetry in the $1$-direction on $T^4$ requires that 
\bea \label{n-symmetry}
k & = & k_1 \cr
k_2 & = & k_{12} \cr
k_3 & = & k_{13} \cr
k_4 & = & k_{14} \cr
k_{23} & = & k_{123} \cr
k_{24} & = & k_{124} \cr
k_{34} & = & k_{134} \cr
k_{234} & = & k_{1234} 
\eea
in addition to (\ref{n-equations}). When $m$ is such that $[\tilde{m}] = 0$, each of the $15$  a priori non-trivial transformations may obey such a symmetry requirement. When $[\tilde{m}] = 1$ this applies to only $3$ of them, and when $[\tilde{m}] = 2$ to none. For a given symmetry compatible with a certain $m$, there are now $16$  solutions to the equations (\ref{n-equations}) and (\ref{n-symmetry}) characterized by the number $2 d$ of multiplicities that are equal to $1$ modulo $2$: 
\beq
\begin{array}{rrrrrrrrrr}
& 2 d = 0 & 2 & 4 & 6 & 8 & 10 & 12 & 14 & 16 \cr
\hline
[\tilde{m}] = 0 & 1 & & & & 14 & & & & 1 \cr
1 & & & 4 & & 8 & & 4 & & \cr
2 & & & & & & & & & \cr
\end{array}
\eeq
Furthermore, the quantum mechanical states in the $V_s$ spaces associated with the points of $\Z_2^4$  must be pairwise equal for the action to have a fixed point. The generating function for the number of these invariant states (which are a subset of the total number of states computed above) is thus given by
\beq
R (q, y) \Po^d (q^2) \Pe^{8 - d} (q^2) .
\eeq

Still assuming that large gauge transformations parametrized by $v \in C$ act trivially in all directions on the torus, we may now determine the number of continua of states for different values of the electric 't~Hooft flux $e \in H^3 (T^4, C)$. By assumption, $e = v \tilde{e}$ for some $\tilde{e} \in H^3 (T^4, \Z_2)$. If all large gauge transformations parametrized by $s \in C$ (or equivalently by $c \in C$) act freely, there will be a fraction $\frac{1}{16}$ of the total number for each of the $16$ possible values of $\tilde{e}$. But if a certain large gauge transformation has a fixed point, the corresponding component of $\tilde{e}$ must be $0$. There will then be a fraction $\frac{1}{8} = \frac{1}{16} + \frac{1}{16}$ of the total number for each of the remaining $8$ possible values of $\tilde{e}$.

We begin with $[\tilde{m}] = 0$, i.e. $[m] = 0$ for $n$ odd and $[m] = 000$ for $n$ even. Taking $\tilde{e} = 0$ gives $[f] = 0$ and $[f] = 000$ respectively. It is convenient to add the corresponding generating functions $Z_0$ and $Z_{000}$ (which are easily distinguished, since they only contain powers of the form $q^{4 k + 2}$ and $q^{4 k}$ respectively). Since there are $15$ large gauge transformations that may have a fixed point, we get
\bea
Z_0 + Z_{000} & = & \frac{1}{16} R(q, y) \left(\Pe^{16} (q) + 30 \Po^8 (q) \Pe^8 (q) + \Po^{16} (q) \right)  \cr
& & + \frac{15}{16} R(q, y) \left(\Pe^8 (q^2) + 14 \Po^4 (q^2) \Pe^4 (q^2) + \Po^8 (q^2) \right) .
\eea
For each of the $15$ non-zero values of $\tilde{e}$, we have $[f] = 0^\prime$ and $[f] = 110$ for $n$ odd and $n$ even respectively. Taking into account that the total number of states is 
\beq
R(q, y) \left(\Pe^{16} (q) + 30 \Pe^8 (q) \Po^8 (q) + \Po^{16} (q) \right)
\eeq
we see that the sum of the corresponding generating functions is
\bea
Z_{0^\prime} + Z_{110} & = & \frac{1}{16} R(q, y) \left(\Pe^{16} (q) + 30 \Po^8 (q) \Pe^8 (q) + \Po^{16} (q) \right)  \cr
& & - \frac{1}{16} R(q, y) \left(\Pe^8 (q^2) + 14 \Po^4 (q^2) \Pe^4 (q^2) + \Po^8 (q^2) \right) . \cr
& & 
\eea

Next, we consider $[\tilde{m}] = 1$, i.e. $[m] = 0^\prime$ for $n$ odd and $[m] = 110$ for $n$ even. There are $4$ values of $\tilde{e}$ that give $[f] = 0^\prime$ and $[f] = 110$, while the remaining $12$ give $[f] = 0^{\prime \prime}$ and $[f] = 220$ for $n$ odd and $n$ even respectively.  The total number of states is
\beq
R(q, y) (4 \Pe^{12} (q) \Po^4 (q) + 24 \Pe^8 (q) \Po^8 (q) + 4 \Pe^4 (q) \Po^8 (q)) ,
\eeq
and there are $3$ large gauge transformations that may have a fixed point. We thus get
\bea
Z_{0^\prime} + Z_{110} & = & \frac{1}{16} R(q, y) \left(4 \Pe^{12} (q) \Po^4 (q) + 24 \Pe^8 (q) \Po^8 (q) + 4 \Pe^4 (q) \Po^8 (q) \right) \cr
& & + \frac{3}{16} R(q, y) \left(4 \Pe^6 (q^2) \Po^2 (q^2) \right. \cr
& & \left. + 8 \Pe^4 (q^2) \Po^4 (q^2) + 4 \Pe^2 (q^2) \Po^6 (q^2) \right) 
\eea
and
\bea
Z_{0^{\prime \prime}} + Z_{220} & = & \frac{1}{16} R(q, y) \left(4 \Pe^{12} (q) \Po^4 (q) + 24 \Pe^8 (q) \Po^8 (q) + 4 \Pe^4 (q) \Po^8 (q) \right) \cr
& & - \frac{1}{16} R(q, y) \left(4 \Pe^6 (q^2) \Po^2 (q^2) \right. \cr
& & \left. + 8 \Pe^4 (q^2) \Po^4 (q^2) + 4 \Pe^2 (q^2) \Po^6 (q^2) \right)  .
\eea

Finally, we take $[\tilde{m}] = 2$, i.e. $[m] = 0^{\prime \prime}$ for $n$ odd and $[m] = 220$ for $n$ even. For all $\tilde{e}$, we get $[f] = 0^{\prime \prime}$ and $[f] = 220$ respectively. The generating functions are given by
\bea
Z_{0^{\prime \prime}} + Z_{220}  & = & \frac{1}{16} R (q, y) \left(16 \Pe^{10} (q) \Po^6 (q) + 16 \Pe^6 (q) \Po^{10} (q) \right) .
\eea

\subsubsection{Electric 't~Hooft flux without vector structure}
We must now consider the possibility that different liftings of the holonomy from $\SO (4 k + 2)$ to $\Spin (4 k + 2)$, i.e. large gauge transformations parametrized by $v \in C$, give rise to different states. The corresponding component of $e \in H^3 (T^4, C)$ will then equal $s$ or $c$. We say that such $e$ have no vector structure.

This will happen if the $8$ multiplicities  associated with the points of a codimension $1$ hyperplane of $\Z_2^4$ are zero. For a large gauge transformation in e.g. the $4$-direction on $T^4$, this means that either
\beq \label{n-empty}
k = k_1 = k_2 = k_3 = k_{12} = k_{13} = k_{23} = k_{123} = 0 
\eeq
or 
\beq \label{n-empty2}
k_4 = k_{14} = k_{24} = k_{34} = k_{124} = k_{134} = k_{234} = k_{1234} = 0 
\eeq
in addition to (\ref{n-equations}). (This choice of direction is only possible if $\tilde{m}_{14} = \tilde{m}_{24} = \tilde{m}_{34} = 0$.) Indeed, if a point from each set of $8$ is occupied, then conjugation by a $\frac{\pi}{2}$ rotation in a suitable $2$-plane will interchange the two liftings, according to the gamma-matrix formula
\beq
\exp (\frac{\pi}{2} \gamma_1 \gamma_2) \gamma_2 \exp(-\frac{\pi}{2} \gamma_1 \gamma_2) = - \gamma_2 .
\eeq
Also, the quartets of angles $(\theta_1, \theta_2, \theta_3, \theta_4)$ at generic values are constrained to be pairwise equal, since the lifting of the holonomy is otherwise equivalent to shifting an angle by $2 \pi$. The two sets of $8$ equations are related by a large gauge transformation parametrized by $v \in C$ in the chosen direction, i.e. the $4$-direction in this example. So there will be a fraction $\frac{1}{2}$ of the extra states for each of the values $s$ and $c$ of the corresponding component of $e$. 

Each set of $8$ equations reduces the number of independent multiplicites to $8$. Furthermore, $4$ of the equations in (\ref{n-equations}) are identically satisified, leaving us with $7$ independent equations, so there are $2^{8-7} = 2$ solutions. Again, they are characterized by the number $2 d$ of multiplicities that are equal to $1$ modulo $2$. Depending on the type $[\tilde{m}]$ we have
\beq
\begin{array}{rrrrrrrrrr}
& 2 d = 0 & 2 & 4 & 6 & 8 & 10 & 12 & 14 & 16 \cr
\hline
[\tilde{m}] = 0 & 1 & & & & 1 & & & &  \cr
1 & & & 2 & &  & &  & & \cr
2 & & & & & & & & & \cr
\end{array}
\eeq
Each solution gives rise to a generating function
\beq
R (q^2, y) \Po^{2 d} (q) \Pe^{8 - 2 d} (q) .
\eeq

Finally, we must also consider the possibility of combining the two phenomena described so far: A configuration could be such that large gauge transformations parametrized by $s$ or $c$ act with a fixed point in one direction and also admit non-trivial large gauge transformations parametrized by $v$ in some other direction. This is described by the equations (\ref{n-equations}), (\ref{n-symmetry}), and (\ref{n-empty}) or (\ref{n-empty2}). Depending on the value of $\tilde{m}$, there are various possibilities. But the number of solutions is still given by
\beq
\begin{array}{rrrrrrrrrr}
& 2 d = 0 & 2 & 4 & 6 & 8 & 10 & 12 & 14 & 16 \cr
\hline
[\tilde{m}] = 0 & 1 & & & & 1 & & & &  \cr
1 & & & 2 & &  & &  & & \cr
2 & & & & & & & & & \cr
\end{array}
\eeq
where, however, each solution now gives rise to a generating function
\beq
R (q^2, y) \Po^{d} (q^2) \Pe^{4 - d} (q^2) .
\eeq

For $n$ odd, low-energy states for which $e \in H^3 (T^4, \Z_4)$ has no vector structure always have a value of $f$ in the orbit $[f] = 1$. For $n$ even, we have to consider the different values of $[\tilde{m}]$ separately: 

For $[\tilde{m}] = 0$, we get $[f] = 011$ or $[f] = 101$ if the remaining components of $e$ are $0$ and $[f] = 111$ otherwise. We thus get
\bea
Z_{1} + Z_{011} & = & Z_{1} + Z_{101} \cr
& = & \frac{1}{8} R (q^2, y) \left(\Pe^8 (q) + \Po^8 (q) \right) \cr
& & + \frac{7}{8} R (q^2, y) \left(\Pe^4 (q^2) + \Po^4 (q^2) \right) .
\eea
and
\bea
Z_{1} + Z_{111} & = & \frac{1}{8} R (q^2, y) \left(\Pe^8 (q) + \Po^8 (q) \right) \cr
& & - \frac{1}{8} R (q^2, y) \left(\Pe^4 (q^2) + \Po^4 (q^2) \right) ,
\eea
since there are $7$ large gauge transformations parametrized by $s \in C$ or $c \in C$ that may act with a fixed point. 

For $[\tilde{m}] = 1$, we get $[f] = 111$ if the remaining components of $e$ are $0$ and $[f] = 221$ otherwise:
 \bea
Z_{1} + Z_{111} & = & \frac{1}{4} R (q^2, y) \Pe^4 (q) \Po^4 (q) \cr
& & + \frac{3}{4} R (q^2, y) \Pe^2 (q^2) \Po^2 (q^2) .
\eea
and
\bea
Z_{1} + Z_{221} & = & \frac{1}{4} R (q^2, y) \Pe^4 (q) \Po^4 (q) \cr
& & - \frac{1}{4} R (q^2, y) \Pe^2 (q^2) \Po^2 (q^2) ,
\eea
since there are $3$ large gauge transformations parametrized by $s \in C$ or $c \in C$ that may act with a fixed point. 

For $[\tilde{m}] = 2$, there are no low-energy states for which $e$ has no vector structure. 

\subsection{Bundles with half vector structure}
As described above, these bundles have $m \mod v = m^v \in H^2 (T^4, \Z_2)$ such that $[m^v] = 1$. We may e.g. take
\beq \label{m^v-example}
m^v_{ij} = \left( \begin{array}{cccc}
0 & 1 & 0 & 0 \cr
1 & 0 & 0 & 0 \cr
0 & 0 & 0 & 0 \cr
0 & 0 & 0 & 0
\end{array}
\right) .
\eeq
The corresponding orbits are
\beq
[m] = 1, (1^{\prime})
\eeq
for $n$ odd, and
\beq
[m] = 011, 101,111, 221, (211, 121)
\eeq
for $n$ even, where the $3$ orbits in parentheses have empty low-energy spectra.

The holonomies may be conjugated to a subgroup
\bea
\Spin (2 n) & \supset & \Spin (2 n - 4 s) \times \Spin (4) \times \ldots \times \Spin (4) / \sim \cr
& \simeq & \Spin (2 n - 4 s) \times \SU (2) \times \SU (2) \times \ldots \times \SU (2) \times \SU (2) / \sim , \cr
& & 
\eea
for certain values of $s$, $0 \leq s \leq \frac{n}{2}$. The equivalence relation $\sim$ identifies the center element $v$ of $\Spin (2 n - 4 s)$ with the center element $v = (-\id_2, - \id_2)$ of one of the $s$ $\Spin (4) \simeq \SU (2) \times \SU (2)$ factors. The decomposition into irreducible terms of the adjoint representation of $\Spin (2 n)$ under this subgroup is
\bea
\frac{1}{2} 2 n (2 n - 1) & = & \frac{1}{2} (2 n - 4 s) (2 n - 4 s - 1) \cr
& & \oplus \bigoplus_1^s 6 \cr
& & \oplus \bigoplus_1^{\frac{1}{2} s (s - 1)} (4, 4) \cr
& & \oplus \bigoplus_1^s (2 n - 4 s, 4) \cr
& = & \frac{1}{2} (2 n - 4 s) (2 n - 4 s - 1) \cr
& & \oplus \bigoplus_1^s (3, 1) \oplus (1, 3)  \cr
& & \oplus \bigoplus_1^{\frac{1}{2} s (s - 1)} (2, 2, 2, 2) \cr
& & \oplus \bigoplus_1^s (2 n - 4 s, 2, 2) .
\eea

The holonomies now take the form
\beq
U_i =  \left(V_i, W_i^{(1)}, T_i^{(1)}, \ldots, W_i^{(s)}, T_i^{(s)} \right) .
\eeq
The $V_i  \in \Spin (2 n - 4 s)$ are chosen to obey the almost commutation relations determined by $m$. For $n$ odd, the simplest example is for $s = \frac{1}{2} (n - 3)$, where we can use the isomorphism $\Spin (2 n - 4 s) = \Spin (6) \simeq \SU (4)$ and the construction described for the $\Phi = A_3$ model. For $n$ even, the simplest example is for $s = \frac{1}{2} n$, so that $\Spin (2 n - 4 s) \simeq 1$ is the trivial group. In any case, we will define $s$ so that all $\frac{1}{2} (2 n - 4s) (2 n - 4s - 1)$ generators in the adjoint representation of $\Spin (2 n - 4 s)$ are broken by the $V_i$.

Each of the $s$ quadruples $W_i^{(1)}, \ldots, W_i^{(s)} \in \SU (2)$ is chosen to obey the almost commutation relations determined by $m^v \in H^2 (T^4, \Z_2)$, where $\Z_2$ is identified with the center of $\SU (2)$. This is given by the construction described for the $\Phi = A_1$ model. These  quadruples thus break all generators of the  $(3, 1)$ terms.

Finally, the $s$ quadruples $T_i^{(1)}, \ldots, T_i^{(s)}$ take their values in a maximal torus of $\SU (2)$ generated by e.g. the Pauli matrix $\sigma = \sigma_3$, i.e. each of them is of the form
\beq
\column{T_1}{T_2}{T_3}{T_4} = \column{\exp (i \sigma \theta_1)}{\exp (i \sigma \theta_2)}{\exp (i \sigma \theta_3)}{\exp (i \sigma \theta_4)} 
\eeq
for some quartet of angular variables $(\theta_1, \theta_2, \theta_3, \theta_4)$. With the value of $m^v$ given e.g. by (\ref{m^v-example}), $T_1$ and $T_2$ are defined modulo the non-trivial center element $-\id_2$ of $\SU (2)$. Generically,  these quadruples break the generators of the $(1, 3)$ terms to $u (1)^s$, but for the $4$ particular values
\beq
\column{T_1}{T_2}{T_3}{T_4} =  \column{\id_2}{\id_2}{\id_2}{\id_2} , \column{\id_2}{\id_2}{-\id_2}{\id_2}, \column{\id_2}{\id_2}{\id_2}{-\id_2}, \column{\id_2}{\id_2}{-\id_2}{-\id_2} ,  
\eeq
the corresponding $u (1)$ term is enhanced to $\su (2) \simeq \symp (2)$.

To analyze the $(2, 2, 2, 2)$ terms, we need the spectrum of one of the $\frac{1}{2}s (s - 1)$ pairs $W_i \otimes W_i^\prime$ on the representation $(2, 2)$ of $\SU (2) \times \SU (2)$. E.g. with $m^v$ as in (\ref{m^v-example}), we get
\beq
\column{W_1 \otimes W_1^\prime}{W_2 \otimes W_2^\prime}{W_3 \otimes W_3^\prime}{W_4 \otimes W_4^\prime} = \column{1}{1}{1}{1} , \column{1}{-1}{1}{1} , \column{-1}{1}{1}{1} , \column{-1}{-1}{1}{1} .
\eeq
It follows that if e.g. $l$ quartets $T_i$ are equal (up to signs), there are $2 \cdot \frac{1}{2} l (l - 1)$ unbroken generators, and $u (1)^l$ is enhanced to $u (l) \simeq \su (l) \oplus u (1)$. Furthermore, if these $l$ quartets $T_i$ take one of the four particular values above, there are actually $4 \cdot \frac{1}{2} l (l - 1)$ unbroken generators, and $\symp (2)^l$ is enhanced to $\symp (2 l)$. Finally, if these $l$ quartets $T_i$ take one of the $12$ particular values
\bea \label{special-T}
\column{T_1}{T_2}{T_3}{T_4} & = & \column{\id_2}{i \sigma}{\id_2}{\id_2} , \column{\id_2}{i \sigma}{\id_2}{-\id_2}, \column{\id_2}{i \sigma}{-\id_2}{\id_2}, \column{\id_2}{i \sigma}{-\id_2}{-\id_2}, \cr 
& &
\column{i \sigma}{\id_2}{\id_2}{\id_2} , \column{i \sigma}{\id_2}{\id_2}{-\id_2}, \column{i \sigma}{\id_2}{-\id_2}{\id_2}, \column{i \sigma}{\id_2}{-\id_2}{-\id_2}, \cr 
& &
\column{i \sigma}{i \sigma}{\id_2}{\id_2} , \column{i \sigma}{i \sigma}{\id_2}{-\id_2}, \column{i \sigma}{i \sigma}{-\id_2}{\id_2}, \column{i \sigma}{i \sigma}{-\id_2}{-\id_2} ,
\eea
there are also $4 \cdot \frac{1}{2} l (l - 1)$ unbroken generators, and $u (1)^l$ is enhanced to $\so (2 l)$.

To analyze the $(2 n - 4 s, 2, 2)$ terms, we need the spectrum of one of the $s$ pairs $V_i \otimes W_i$ on the representation $(2 n - 4 s, 2)$ of $\Spin (2 n - 4s) \times \SU (2)$. Consider the $s = \frac{1}{2} (n - 3)$ case for $n$ odd. E.g. with $m^v$ as in (\ref{m^v-example}), there are four inequivalent choices of $V_i$, characterized by a pair
\beq \label{c-choice}
\left(
\begin{array}{r}
c_3 \cr
c_4
\end{array}
\right) 
=  
\left(
\begin{array}{r}
1 \cr
1
\end{array}
\right) ,
\left(
\begin{array}{r}
1 \cr
-1
\end{array}
\right) ,
\left(
\begin{array}{r}
-1 \cr
1
\end{array}
\right) ,
\left(
\begin{array}{r}
-1 \cr
-1
\end{array}
\right) .
\eeq
The corresponding spectra on the representation $(6, 2)$ are
\bea
\column{V_1 \otimes W_1}{V_2 \otimes W_2}{V_3 \otimes W_3}{V_4 \otimes W_4} & = &
\column{1}{i}{c_3}{c_4} , \column{1}{-i}{c_3}{c_4}, \column{i}{1}{c_3}{c_4} , \column{i}{i}{c_3}{c_4} ,\column{i}{-1}{c_3}{c_4} , \column{i}{-i}{c_3}{c_4} ,   \cr
& & \column{-1}{i}{c_3}{c_4} , \column{-1}{-i}{c_3}{c_4}, \column{-i}{1}{c_3}{c_4} , \column{-i}{i}{c_3}{c_4} ,\column{-i}{-1}{c_3}{c_4} , \column{-i}{-i}{c_3}{c_4} . \cr
& & 
\eea
The four pairs in (\ref{c-choice}) correspond to the four columns of special values in (\ref{special-T}). When $l$ quartets are equal to one of the three values in that column, $\so (2 l)$ is enhanced to $\so (2 l + 1)$. 
 
Still for $n$ odd, $4$ other solutions associated with the columns in (\ref{special-T}) have $\so (2 l)$ to $\so (2 l + 1)$ enhancement when $l$ quartets are equal to one the nine values not in that column. $12$ solutions are associated with a row and a column, and have enhancement for the five values that are in that row but not in that column, or not in that row but in that column. Finally, $12$ more solutions have enhancement for the seven values that are in that row and in that column, or not in that row and not in that column.

So for $n$ odd, the number of solutions where $d = n - 2 s$ of the special values (\ref{special-T}) have $\so (2 l)$ enhanced to $\so (2 l + 1)$ is given by the following table:
\beq
\begin{array}{rrrrrrrrrrrrrr}
& d = 0 & 1 & 2 & 3 & 4 & 5 & 6 & 7 & 8 & 9 & 10 & 11 & 12 \cr
\hline
[m] = 1 & & & & 4 & & 12 & & 12 & & 4 & & & \cr
\end{array}
\eeq

The case $n$ even may be analyzed in an analogous way: For $[m]Ê= 011$ or $[m] = 101$, one solution has no enhancement at any value, and one solution has enhancement at all twelve values. Three solutions are associated with the rows, and have enhancement for the four values in that row, and three more solutions have enhancement for the eight values not in that row. Six solutions are associated with pairs of columns, and have enhancement for the six values in either of these columns. Finally, eighteen solutions are associated with a row and a pair of columns, and have enhancement for the six values that are in that row and in either of these columns, or not in that row and in neither of these columns. For $[m] = 111$, or $[m] = 221$, the value of $m$ distinguishes between the row and the columns, so the pattern is slightly more complicated. The results are summarized in the following table:
\beq
\begin{array}{rrrrrrrrrrrrrr}
& d = 0 & 1 & 2 & 3 & 4 & 5 & 6 & 7 & 8 & 9 & 10 & 11 & 12 \cr
\hline
011 & 1 & & & & 3 & & 24 & & 3 & & & & 1 \cr
101 & 1 & & & & 3 & & 24 & & 3 & & & & 1 \cr
111 & & & 2 & & 4 & & 20 & & 4 & & 2 & & \cr
221 & & & & & 12 & & 8 & & 12 & & & & 
\end{array}
\eeq

In all cases, the generating function for the total number of continua of states associated with such a solution is
\beq
R (q^2, y) Q (q^2)^4 \Po^d (q^2) \Pe^{12 - d} (q^2) .
\eeq

\subsubsection{Electric 't~Hooft flux with vector structure}
With $m^v$ as in (\ref{m^v-example}), large gauge transformations parametrized by $v \in C$ act trivially in the $1$- and $2$-directions. In e.g. the $3$-direction, it acts freely, unless the configuration is symmetric under exchange of the first and the third columns and the second and the fourth columns in (\ref{special-T}). This cannot happen for $n$ odd, so there we get
\bea
Z_1 & = & \frac{1}{4} R (q^2, y) Q (q^2)^4 \left(4 \Po^3 (q^2) \Pe^9 (q^2) + 12 \Po^5 (q^2) \Pe^7 (q^2) \right. \cr
& & \left. + 12 \Po^7 (q^2) \Pe^5 (q^2) + 4 \Po^9 (q^2) \Pe^3 (q^2) \right) .
\eea
For $n$ even, there are the following number of possibilities for this to happen:
\beq
\begin{array}{rrrrrrrrrrrrrr}
& d = 0 & 1 & 2 & 3 & 4 & 5 & 6 & 7 & 8 & 9 & 10 & 11 & 12 \cr
\hline
011 & 1 & & & & 3 & & 8 & & 3 & & & & 1 \cr
101 & 1 & & & & 3 & & 8 & & 3 & & & & 1 \cr
111 & & & 2 & & 4 & & 4 & & 4 & & 2 & & \cr
221 & & & & & & & & & & & & & 
\end{array}
\eeq
The generating function for the total number of continua of states associated with such a solution is
\beq
R (q^2, y) Q (q^4)^2 \Po^{d/2} (q^4) \Pe^{6 - d/2} (q^4) .
\eeq
For $[m] = 011$ or $[m] = 101$, we thus get
\bea
Z_{011} & = & Z_{101} \cr
& = &  \frac{1}{4} R(q^2, y) Q^4 (q^2) \left( \Pe^{12} (q^2) + 3 \Pe^8 (q^2) \Po^4 (q^2) \right. \cr
& & \left. + 24 \Pe^6 (q^2) \Po^6 (q^2) + 3 \Pe^4 (q^2) \Po^8 (q^2) + \Po^{12} (q^2) \right) \cr
& & + \frac{3}{4} R (q^2, y) Q^2 (q^4) \left( \Pe^6 (q^4) + 3 \Pe^4 (q^4) \Po^2 (q^4) \right. \cr
& & \left. + 8 \Pe^3 (q^4) \Po^3 (q^4) + 3 \Pe^2 (q^4) \Po^4 (q^4) + \Po^6 (q^4) \right) \cr
Z_{111} & = & \frac{1}{4} R (q^2, y) Q^4 (q^2) \left( \Pe^{12} (q^2) + 3 \Pe^8 (q^2) \Po^4 (q^2) \right. \cr
& & \left. + 24 \Pe^6 (q^2) \Po^6 (q^2) + 3 \Pe^4 (q^2) \Po^8 (q^2) + \Po^{12} (q^2) \right) \cr
& & - \frac{1}{4} R (q^2, y) Q^2 (q^4) \left( \Pe^6 (q^4) + 3 \Pe^4 (q^4) \Po^2 (q^4) \right. \cr
& & \left. + 8 \Pe^3 (q^4) \Po^3 (q^4) + 3 \Pe^2 (q^4) \Po^4 (q^4) + \Po^6 (q^4) \right) .
\eea
whereas for $[m] = 111$
\bea
Z_{111} & = & \frac{1}{4} R (q^2, y) Q^4 (q^2) \left(2 \Pe^{10} (q^2) \Po^2 (q^2) + 4 \Pe^8 (q^2) \Po^4 (q^2) \right. \cr
& & \left. + 20 \Pe^6 (q^2) \Po^6 (q^2) + 4 \Pe^4 (q^2) \Po^8 (q^2) + 2 \Pe^2 (q^2) \Po^{10} (q^2) \right) \cr
& & + \frac{1}{4} R (q^2, y) Q^2 (q^4) \left( 2 \Pe^5 (q^4) \Po (q^4) + 4 \Pe^4 (q^4) \Po^2 (q^4) \right. \cr
& & \left. + 4 \Pe^3 (q^4) \Po^3 (q^4) + 4 \Pe^2 (q^4) \Po^4 (q^4) + 2 \Pe (q^4) \Po^5 (q^4) \right) \cr
Z_{221} & = & \frac{1}{4} R (q^2, y) Q^4 (q^2) \left(2 \Pe^{10} (q^2) \Po^2 (q^2) + 4 \Pe^8 (q^2) \Po^4 (q^2) \right. \cr
& & \left. + 20 \Pe^6 (q^2) \Po^6 (q^2) + 4 \Pe^4 (q^2) \Po^8 (q^2) + 2 \Pe^2 (q^2) \Po^{10} (q^2) \right) \cr
& & - \frac{1}{4} R (q^2, y) Q^2 (q^4) \left( 2 \Pe^5 (q^4) \Po (q^4) + 4 \Pe^4 (q^4) \Po^2 (q^4) \right. \cr
& & \left. + 4 \Pe^3 (q^4) \Po^3 (q^4) + 4 \Pe^2 (q^4) \Po^4 (q^4) + 2 \Pe (q^4) \Po^5 (q^4) \right) .
\eea
Finally, for $[m] = 221$
\bea
Z_{221} & = & \frac{1}{4} R(q^2, y) Q^4 (q^2) \left(12 \Po^4 (q^2) \Pe^8 (q^2) \right. \cr
& & \left. + 8 \Po^6 (q^2) \Pe^6 (q^2) + 12 \Po^8 (q^2) \Pe^4 (q^2) \right) . \cr
& & 
\eea

\subsubsection{Electric 't~Hooft flux without vector structure}
With $m^v$ as in (\ref{m^v-example}), a large gauge transformation parametrized by $s \in C$  or $c \in C$ acts trivially in the $3$- and $4$-directions. In e.g. the $1$-direction, it acts non-trivially only if the last two rows in (\ref{special-T}) are empty. This cannot happen for $n$ odd. For $n$ even, we have the following possibilities:
\beq
\begin{array}{rrrrrrrrrrrrrr}
& d = 0 & 1 & 2 & 3 & 4 & 5 & 6 & 7 & 8 & 9 & 10 & 11 & 12 \cr
\hline
011 & 1 & & & & 1 & & & & & & & & \cr
101 & 1 & & & & 1 & & & & & & & & \cr
111 & & & 2 & & & & & & & & & & \cr
221 & & & & & & & & & & & & & 
\end{array}
\eeq
The total number of continua of states for such a solution is given by
\beq
R (q^4, y) Q^4 (q^2) \Po^d (q^2) \Pe^{4 - d} (q^2) .
\eeq 
Imposing symmetry in e.g. the $3$-direction still gives
\beq
\begin{array}{rrrrrrrrrrrrrr}
& d = 0 & 1 & 2 & 3 & 4 & 5 & 6 & 7 & 8 & 9 & 10 & 11 & 12 \cr
\hline
011 & 1 & & & & 1 & & & & & & & & \cr
101 & 1 & & & & 1 & & & & & & & & \cr
111 & & & 2 & & & & & & & & & & \cr
221 & & & & & & & & & & & & & 
\end{array}
\eeq
 but with the number of states now given by
\beq
R (q^4, y) Q^2 (q^4) \Po^{d/2} (q^4) \Pe^{2 - d/2} (q^4) .
\eeq
For $[m] = 011$ or $[m]Ê= 101$, we get
\bea
Z_{112} + Z_{022} & = & Z_{112} + Z_{202} \cr
& = & \frac{1}{4} R (q^4, y) Q^4 (q^2) \left(\Pe^4 (q^2) + \Po^4 (q^2) \right) \cr
& & + \frac{3}{4} R (q^4, y) Q^2 (q^4) \left(\Pe^2 (q^4) + \Po^2 (q^4) \right) \cr
Z_{112} + Z_{122} & = & Z_{112} + Z_{212} \cr
& = & \frac{1}{4} R (q^4, y) Q^4 (q^2) \left(\Pe^4 (q^2) + \Po^4 (q^2) \right) \cr
& & - \frac{1}{4} R (q^4, y) Q^2 (q^4) \left(\Pe^2 (q^4) + \Po^2 (q^4) \right) .
\eea
For $[m] = 111$, we get
\bea
Z_{112} + Z_{122} & = & Z_{112} + Z_{212} \cr
& = & \frac{1}{4} R (q^4, y) Q^4 (q^2) 2 \Po^2 (q^2) \Pe^2 (q^2) \cr
& & + \frac{1}{4} R ^(q^4, y) Q^2 (q^4) 2 \Po (q^4) \Pe (q^4) \cr
Z_{222} + Z_{122} & = & Z_{222} + Z_{212} \cr
& = & \frac{1}{4} R (q^4, y) Q^4 (q^2) 2 \Po^2 (q^2) \Pe^2 (q^2) \cr
& & - \frac{1}{4} R ^(q^4, y) Q^2 (q^4) 2 \Po (q^4) \Pe (q^4) .
\eea
Finally, for $[m] = 221$, this does not happen.

\subsection{Bundles with no vector structure}
As described above, these bundles have $m \mod v = m^v \in H^2 (T^4, \Z_2)$ such that $[m^v] = 2$. For $n$ odd, no such bundles admit low-energy states. For $n$ even, the possibilities are 
\beq
[m] = 022, 202, 112, 122,  212, 222 .
\eeq

Possible unbroken subalgebras of $\so (2 n)$ are of the form
\bea
h & \simeq & \symp (2 k_1) \oplus \ldots \oplus \symp (2 k_6) \cr
& & \oplus \so (n_1) \oplus \ldots \oplus \so (n_{10}) \cr
& & \oplus \su (l_1) \oplus \ldots \oplus \su (l_{r+1}) \cr
& & \oplus u(1)^r ,
\eea
such that the rank of $h$ equals $\frac{n}{4}$ for $n = 0 \mod 4$ and $\frac{n- 2}{4}$ for $n = 2 \mod 4$. Furthermore, a number $0 \leq d \leq 10$ of $n_1, \ldots, n_{10}$ are odd and the remaining $10 - d$ are even. Depending on $[m]$ and $d$, there are the following number of solutions:
\beq
\begin{array}{rrrrrrrrrrrr}
& d = 0 & 1 & 2 & 3 & 4 & 5 & 6 & 7 & 8 & 9 & 10 \cr
\hline
[m] = 022 & 1 & & & & 15 & & 15 & & & & 1 \cr
202 & 1 & & & & 15 & & 15 & & & & 1 \cr
112 & & 1 & & 6 & & 18 & & 6 & & 1 & \cr
122 & & & 3 & & 13 & & 13 & & 3 & & \cr
212 & & & 3 & & 13 & & 13 & & 3 & & \cr
222 & & & & 10 & & 12 & & 10 & & & \cr
\end{array}
\eeq
The total number of continua of states for such a solution is given by
\beq
R (q^4, y) Q^6 (q^4) \Po^d (q^4) \Pe^{10-d} (q^4) .
\eeq
Large gauge transformations parametrized by $v \in C$ always act trivially, and those parametrized by $s \in C$ or equivalently $c \in C$ always act freely. We thus get the following generating functions:
\bea
Z_{022} & = & Z_{202} \cr
& = & R (q^4, y) Q^6 (q^4) \left( \Pe^{10} (q^4) + 15 \Po^4 (q^4) \Pe^6 (q^4) \right. \cr
& & \left. + 15 \Pe^6 (q^4) \Po^4 (q^4) + \Po^{10} (q^4) \right) \cr
Z_{112} & = & R (q^4, y) Q^6 (q^4) \left( \Po (q^4) \Pe^9 (q^4) + 6 \Po^3 (q^4) \Pe^7 (q^4) \right. \cr
& & \left. + 18 \Pe^5 (q^4) \Po^5 (q^4) + 6 \Po^7 (q^4) \Pe^3 (q^4) + \Po^9 (q^4) \Pe (q^4) \right) \cr
Z_{122} & = & Z_{212} \cr
& & R (q^4, y) Q^6 (q^4) \left( 3 \Po^2 (q^4) \Pe^8 (q^4) + 13 \Po^4 (q^4) \Pe^6 (q^4) \right. \cr
& & \left. + 13 \Pe^6 (q^4) \Po^4 (q^4) + 3 \Po^8 (q^4) \Pe^2 (q^4) \right) \cr
Z_{222} & = & R (q^4, y) Q^6 (q^4) \left( 10 \Po^3 (q^4) \Pe^7 (q^4) \right. \cr
& & \left. + 12 \Pe^5 (q^4) \Po^5 (q^4) + 10 \Po^7 (q^4) \Pe^3 (q^4) \right) .
\eea

\subsection{The results}
The number of rank $r$ continua of 't~Hooft flux $f \in H^3 (T^5, C)$ in the $\Phi = D_n$ theory is given by the coefficient of $y^r q^{2n}$ in the power series expansion of the generating function $Z_f (q, y)$. We have computed alternative expressions for these functions for $f$ with vector structure, half vector structure, and no vector structure, and found that they are given by $R (q, y)$, $R (q^2, y)$, and $R (q^4, y)$ respectively, times a polynomial in the functions $\Pe$ and $\Po$ with various arguments. The equality of these different expressions for a single $\SL_5 (\Z)$ orbit $[f]$ amounts to certain combinatorial identities for $\Pe$ and $\Po$, that may be proven e.g. by relating them to modular forms. (See e.g. \cite{Farkas-Kra}.) Here, however, we will only expand the expressions for $Z_{[f]} (q, y)$ to the first few orders in $q$.

For $n$ odd, one finds for $f \in H^3 (T^5, \Z_4)$ with vector structure 
\bea
Z_{0} & = & y q^2 \cr
& & + (1 + 2 y + y^2 + y^3) q^6 \cr
& & + (32 + 35 y + 4 y^2 + 3 y^3 + y^4 + y^5) q^{10} \cr
& & + (528 + 285 y + 71 y^2 + 39 y^3 + 5 y^4 + 3 y^5 + y^6 + y^7) q^{14} + \ldots \cr
Z_{0^\prime} & = & (1 + y) q^6 \cr
& & + (32 + 12 y + 2 y^2 + y^3) q^{10} \cr
& & + (528 + 198 y + 46 y^2 + 13 y^3 + 2 y^4 + y^5) q^{14} \cr
& & + (6016 + 2626 y + 772 y^2 + 213 y^3 + 47 y^4 + 13 y^5 + 2 y^6 + y^7) q^{18} + \ldots \cr
Z_{0^{\prime \prime}} & = & q^6 \cr
& & + (32 + 7 y + y^2) q^{10} \cr
& & + (528 + 175 y + 40 y^2 + 7 y^3 + y^4) q^{14} \cr
& & + (6016 + 2547 y + 743 y^2 + 183 y^3 + 40 y^4 + 7 y^5 + y^6) q^{18} + \ldots ,
\eea
for $f$ with half vector structure
\bea
Z_{1} & = & q^6 \cr
& & + (10 + y) q^{10} \cr
& & + (67 + 11 y + y^2) q^{14} \cr
& & + (350 + 78 y + 11 y^2 + y^3) q^{18} + \ldots \cr
Z_{1^\prime} & = & 0 ,
\eea
and for $f$ with no vector structure
\bea
Z_{2} & = &  0 .
\eea
For $n$ even, one finds for $f \in H^3 (T^5, \Z_2 \times \Z_2)$ with vector structure
\bea
Z_{000} & = & 1 \cr
& & + (1 + y + y^2) q^4 \cr
& & + (32 + 3 y + 3 y^2 + y^3 + y^4) q^8 \cr
& & + (218 + 67 y + 38 y^2 + 5 y^3 + 3 y^4 + y^5 + y^6) q^{12} + \ldots \cr
Z_{110} & = & q^4 \cr
& & + (10 + 2 y + y^2) q^8 \cr
& & + (154 + 44 y + 13 y^2 + 2 y^3 + y^4) q^{12} \cr
& & + (1900 + 726 y + 211 y^2 + 47 y^3 + 13 y^4 + 2 y^5 + y^6) q^{16} + \ldots \cr
Z_{220} & = & (6 + y) q^8 \cr
& & + (136 + 39 y + 7 y^2 + y^3) q^{12} \cr
& & + (1844 + 703 y + 182 y^2 + 40 y^3 + 7 y^4 + y^5) q^{16} \cr
& & + (18384 + 8563 y + 2729 y^2 + 750 y^3 + 183 y^4 + 40 y^5 + 7 y^6 + y^7) q^{20} + \ldots ,
\eea
for $f$ with half vector structure
\bea
Z_{011} & = & Z_{101} \cr
& = & 1 \cr
& & + (1 + y) q^4 \cr
& & + (10 + 2 y + y^2) q^8 \cr
& & + (39 + 12 y + 2 y^2 + y^3) q^{12} + \ldots \cr
Z_{111} & = & q^4 \cr
& & + (5 + y) q^8 \cr
& & + (31 + 6 y + y^2) q^{12} \cr
& & + (165 + 37 y + 6 y^2 + y^3) q^{16} + \ldots \cr
Z_{221} & = & 3 q^8 \cr
& & + (26 + 3 y) q^{12} \cr
& & + (155 + 29 y + 3 y^2) q^{16} \cr
& & + (746 + 184 y + 29 y^2 + 3 y^3) q^{20} + \ldots \cr
Z_{211} & = & Z_{121} \cr
& = & 0 ,
\eea
and for $f$ with no vector structure
\bea
Z_{022} & = & Z_{202} \cr
& = & 1 \cr
& & + (6 + y) q^8 \cr
& & + (46 + 7 y + y^2) q^{16} \cr
& & + (297 + 53 y + 7 y^2 + y^3) q^{24} + \ldots \cr
Z_{112} & = & q^4 \cr
& & + (13 + y) q^{12} \cr
& & + (109 + 14 y + y^2) q^{20} \cr
& & + (678 + 123 y + 14 y^2 + y^3) q^{28} + \ldots \cr
Z_{122} & = & Z_{212} \cr
& = & 3 q^8 \cr
& & + (37 + 3 y) q^{16} \cr
& & + (275 + 40 y + 3 y^2) q^{24} \cr
& & + (1546 + 315 y + 40 y^2 + 3 y^3) q^{32} + \ldots \cr
Z_{222} & = & 10 q^{12} \cr
& & + (102 + 10 y) q^{20} \cr
& & + (662 + 112 y + 10 y^2) q^{28} \cr
& & + (3392 + 774 y + 112 y^2 + 10 y^3) q^{36} + \ldots \cr
Z_{122^\prime} & = & Z_{212^\prime} \cr
& = & 0 \cr
Z_{221^\prime} & = & 0 \cr
Z_{222^\prime} & = & 0.
\eea

As a last minor check, we note that for the $\Phi = D_4$ model only, there is an $S_3$ triality group of automorphisms which permute the non-trivial center elements $s, c, v \in C \simeq \Z_2 \times \Z_2$: Restricting attention to the $q^8$ terms, we have 
\bea
Z_{110} & = & Z_{011} = Z_{101} = (10 + 2 y + y^2) q^8 \mod q^{2 n}, n \neq 4 \cr
Z_{220} & = & Z_{022} = Z_{202} = (6 + y) q^8 \cr
Z_{221} & = & Z_{122} = Z_{212} = 3 q^8 .
\eea

\vspace*{10mm} \noindent
{\bf \large Acknowledgements}

This research was supported by grants from the G\"oran Gustafsson foundation and the Swedish Research Council.

I am grateful to Edward Witten for suggesting this approach to the $(2, 0)$ theories, and to Niclas Wyllard for many helpful comments.

 \end{document}